\documentclass[twocolumn, usenatbib]{aastex631}

\usepackage{amsmath,amssymb}
\usepackage{subfigure}
\usepackage{multirow,verbatim}
\newcommand{\etaeff}{\eta_{\rm eff}}

\begin{document}

\title{Effective resistivity in relativistic reconnection: \\a  prescription based on fully kinetic simulations}

\author[0000-0002-6437-5229]{Abigail Moran}
\email{abigail.moran@columbia.edu}
\affiliation{Department of Astronomy, Columbia University,  New York, NY 10027, USA}

\author[0000-0002-1227-2754]{Lorenzo Sironi}
\affiliation{Department of Astronomy, Columbia University,  New York, NY 10027, USA}
\affiliation{Center for Computational Astrophysics, Flatiron Institute, 162 Fifth Ave, New York, NY 10010, USA}

\author[0000-0001-7307-632X]{Aviad Levis}
\affiliation{Department of Computer Science, University of Toronto, Toronto, ON M5S 2E4, Canada}
\affiliation{David A. Dunlap Department of Astronomy \& Astrophysics, University of Toronto, Toronto, ON M5S 3H4, Canada}

\author[0000-0002-7301-3908]{Bart Ripperda}
\affiliation{Canadian Institute for Theoretical Astrophysics, 60 St. George St, Toronto, ON M5S 3H8, Canada}
\affiliation{Department of Physics, University of Toronto, 60 St. George St, Toronto, ON M5S 1A7, Canada}
\affiliation{David A. Dunlap Department of Astronomy \& Astrophysics, University of Toronto, Toronto, ON M5S 3H4, Canada}
\affiliation{Perimeter Institute for Theoretical Physics, 31 Caroline St. North, Waterloo, ON N2L 2Y5, Canada}

\author[0000-0002-0491-1210]{Elias R. Most}
\affiliation{TAPIR, Mailcode 350-17, California Institute of Technology, Pasadena, CA 91125, USA}
\affiliation{Walter Burke Institute for Theoretical Physics, California Institute of Technology, Pasadena, CA 91125, USA}

\author[0000-0001-9508-1234]{Sebastiaan Selvi}
\affiliation{Department of Astronomy, Columbia University,  New York, NY 10027, USA}

\begin{abstract}

A variety of high-energy astrophysical phenomena are powered by the release---via magnetic reconnection---of the energy stored in oppositely directed fields. Single-fluid resistive magnetohydrodynamic (MHD) simulations
with uniform resistivity yield dissipation rates
that are much lower (by
nearly one order of magnitude) than equivalent kinetic
calculations. Reconnection-driven phenomena could be accordingly modeled in resistive MHD employing a non-uniform, ``effective'' resistivity informed by kinetic calculations. 
In this work, we analyze a suite of fully kinetic particle-in-cell (PIC) simulations of relativistic pair-plasma reconnection---where the magnetic energy is greater than the rest mass energy---for different strengths of the guide field orthogonal to the alternating component. We extract an empirical prescription for the effective resistivity, $\eta_{\mathrm{eff}} = \alpha B_0 \mathbf{|J|}^p / \left(|\mathbf{J}|^{p+1}+\left(e n_t c\right)^{p+1}\right)$, where $B_0$ is the reconnecting magnetic field strength, $\bf J$ is the current density, $n_t$ the lab-frame total number density, $e$ the elementary charge, and $c$ the speed of light. The guide field dependence is encoded in $\alpha$ and $p$, which we fit to PIC data. 
This resistivity formulation---which relies only on single-fluid MHD quantities---successfully reproduces the spatial structure and strength of nonideal electric fields, 
and thus provides a promising strategy for enhancing the reconnection rate in resistive MHD simulations.  
\end{abstract}

\section{Introduction} \label{sec:intro}
Strong magnetic fields in astrophysical compact sources provide a reservoir of magnetic energy. This energy can be released to the plasma---resulting in particle acceleration and nonthermal emission---when anti-aligned field lines annihilate in a process called magnetic reconnection. In a number of astrophysical sources, reconnection occurs in the 
relativistic regime, where the magnetic energy exceeds the plasma rest mass energy \citep[for reviews, see] []{hoshino_lyubarsky_12,kagan_15,guo_review_20,guo_review}. Relativistic reconnection can power a variety of high-energy phenomena, such as emission from black hole coronae,
magnetar flares,
blazar jet flares, 
 radio and gamma-ray emission from pulsar magnetospheres, fast radio bursts, and flares from supermassive black hole magnetospheres. 

Magnetic reconnection refers to the breaking and reconnecting of oppositely directed  field lines. This requires the  ``ideal'' condition
\[ 
{\bf E}+\frac{\langle{\bf v}_s\rangle}{c}\times {\bf B}=0
\]
to be violated for each relevant plasma species $s$. Here, ${\bf E}$ and ${\bf B}$ are the electromagnetic fields, while $\langle {\bf v}_s\rangle$ is the mean three-velocity of species $s$. 
In {\it collisional} plasmas, the ideal condition can be broken by resistive effects due to binary particle collisions---encoded by the resistivity appearing in Ohm's law. When resistive effects are not important, the magnetic field is ``frozen'' into the fluid, as prescribed by Alfv\'en's theorem (\citealt{alfven_43}; also known as the flux freezing theorem). In dilute astrophysical plasmas, binary collisions are rare, so the collisional resistivity is often insufficient to break flux freezing on interesting time and length scales. 

Reconnection occurring in the {\it collisionless} regime requires a kinetic description. Since the typical separation between plasma scales and global scales is very large, kinetic descriptions, e.g., employing the particle-in-cell (PIC) method, are unaffordable at realistic scale separations. Fluid-type approaches such as magnetohydrodynamics (MHD), while suitable to model the global dynamics, are by construction collisional, and therefore unable to capture collisionless effects. 
In fact, single-fluid resistive MHD simulations with uniform resistivity yield reconnection rates in the plasmoid-dominated regime that are much lower (by nearly one order of magnitude) than equivalent kinetic calculations \citep{birn_01, cassak_review,Uzdensky, Comisso}. This discrepancy impacts the timescale of reconnection-powered flares, e.g., in black hole magnetospheres \citep{bransgrove_21,Galishnikova}.

A large body of work has focused on identifying the processes that can break the ideal condition in collisionless or weakly collisional plasmas---here, wave-particle interactions provide a form of effective collisionality. In pair plasmas, fast reconnection is mediated by the off-diagonal terms of the pressure tensor \citep{bessho_05,Bessho07,Hesse07,melzani_14a,goodbred_22}, which are also important for electron-ion plasmas in the small, electron-scale diffusion region  \citep{lyons_90,horiuchi_94,cai_97,kuznetsova_98,
egedal_19}. 

By identifying the dominant contributors to the breaking of flux freezing in collisionless plasmas, it may be possible to write the corresponding nonideal electric field as $\eta_{\rm eff} {\bf J}$---here, $\eta_{\rm eff}$ is some effective
resistivity and ${\bf J}$ the electric current density---which could be incorporated in resistive, single-fluid MHD approaches as a kinetically motivated subgrid prescription \citep{Kulsrud,Kulsrud01, Biskamp, Uzdensky03,  Zenitani, Bessho2010, Ripperda2019, Nuno}. In general, parameterizing kinetic effects as an effective resistivity is a nontrivial task (e.g., \citealt{Hirvijoki, Lingam}). By means of a statistical analysis based on PIC simulations, \citet{Selvi} identified the mechanisms driving the nonideal electric field in the generalized Ohm's law, for the case of relativistic pair plasma reconnection. The effective resistivity proposed by \citet{Selvi} for the zero guide field case (and earlier suggested by \citealt{Bessho07,Bessho12}) has been shown to successfully enhance the reconnection rate in resistive MHD simulations \citep{Bugli}.

As we discuss below, the form of effective resistivity proposed by \citet{Selvi} suffers from a few limitations, which may hamper its applicability. In this work, rather than analyzing nonideal terms in the generalized Ohm's law, we adopt an empirical approach. We perform a suite of PIC
simulations of relativistic pair-plasma reconnection with varying guide field strength, and we formulate an empirical prescription for the effective resistivity $\etaeff$, which is derived directly from our PIC runs through a data-driven parameterization. Our proposed model depends only on the electric current density and the plasma number density, both of which are readily available in resistive MHD codes. As compared to \citet{Selvi}, the form of $\etaeff$ that we obtain has {four} main advantages: it is written explicitly in single-fluid MHD quantities, does not depend on spatial derivatives, is coordinate-agnostic, {and is valid for any guide field}. We demonstrate that the formulation of $\etaeff$ we propose successfully reproduces the spatial structure and strength of nonideal electric fields in our PIC simulations, thus providing a
promising strategy for enhancing the reconnection rate in resistive MHD approaches.

\begin{figure*}
    \centering
    \includegraphics[width=0.98\linewidth]{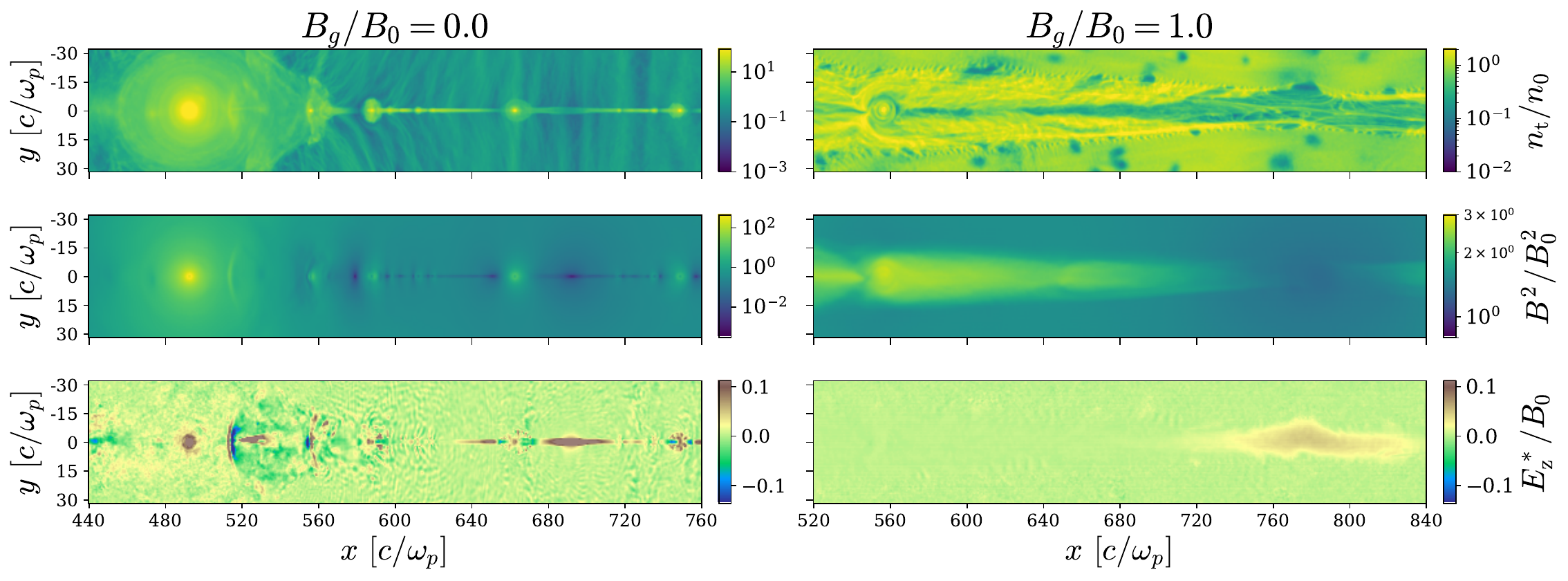}
    \caption{Spatial distribution of the particle number density $n_t$ (top row; in units of $n_0$), of the magnetic energy density (middle row; in units of $B_0^2/8\pi$), and of the $z$-component of the nonideal electric field as defined in \autoref{Eq:eta} (bottom row; in units of $B_0$), for simulations with $B_g/B_0 = 0$ (left) and $B_g/B_0 = 1$ (right). The snapshots are taken at a representative time {to  show the nonideal electric field and plasmoid structure} after the simulations have achieved a quasi-steady state. We define $x=0$ as the edge of the simulation domain {(so, the plot only shows a portion of the domain)}, while $y=0$ in the  midplane.}
    \label{fig:sim_setul}
\end{figure*}

\section{PIC Simulation Setup} \label{sec:PICsim}
Our simulations are performed with the 3D PIC code \texttt{TRISTAN-MP} \citep{Tristan}. We use a 2D $x-y$ domain, but we track all components of the particle velocity and of the electromagnetic fields. Although the physics of particle acceleration in relativistic reconnection is dramatically different between 2D and 3D \citep{zhang_21,zhang_23}, the reconnection rate---which is defined as the plasma inflow velocity and is the focus of our work---is roughly the same \citep[e.g.,][]{sironi_spitkovsky_14,werner_17}.

The in-plane magnetic field is initialized in a ``Harris equilibrium'' \citep{harris_62}, ${\bf B_{\rm in}} = B_0{\bf \hat{x}}\tanh(2\pi y/\Delta)$, where the direction of the in-plane field reverses at $y = 0$ over a thickness $\Delta = 70\,c/\omega_{p}$. Here, $c/\omega_p$ is the depth of the plasma skin, and $\omega_p=\sqrt{4\pi n_0e^2/m}$ the plasma frequency, where $n_0$ is the total number density of electron-positron pairs far from the layer, $m$ the electron / positron mass and $e$ the elementary charge.
We parameterize the field strength in the plane $B_0$ by the magnetization
\[
 \sigma = \frac{B_0^2}{4\pi n_0 m c^2}~,
\]
which we take to be $\sigma=50$. We consider guide fields of magnitude $B_g/B_0 = 0.0, 0.3, 0.6, 1.0$.

The upstream region is initialized with $n_0=64$ particles per cell (including both species). We resolve the plasma skin depth $c/\omega_p$ with five cells and evolve the simulation up to  $4500 \, \omega_p^{-1}$. In \autoref{app:tests}, we choose  $c/\omega_p=20$~cells and demonstrate that our results are robust to spatial resolution. In \autoref{app:tests} we also display cases that include strong synchrotron cooling losses.
For our fiducial runs, the length of the domain in the $x$-direction of plasma outflows is $L_x = 1920\,c/\omega_p$. We use open boundaries for fields and particles along the $x$-direction. The box grows in the $y$-direction as the simulation progresses, allowing for more plasma and magnetic flux to enter the domain. At the end of the simulations, the length of our box along the $y$-axis is comparable to $L_{x}$.

\section{Resistivity Formulation} \label{sec:methods}
We begin by considering the
Ohm's law for resistive relativistic single-fluid MHD \citep{Komissarov}:
\begin{equation}\label{eq: ohm_full}
    \Gamma \left[\mathbf{E}+\frac{\mathbf{v}}{c} \times \mathbf{B} - \frac{1}{c^2}(\mathbf{E}\cdot \mathbf{v})\mathbf{v} \right] = \eta \left(\mathbf{J}-\rho_e \mathbf{v} \right)
\end{equation}
where $\Gamma$ is the bulk fluid Lorentz factor, $\mathbf{v}$ the fluid three-velocity, $\rho_e$ the electric charge density, and $\eta$ the collisional resistivity. In a collisionless plasma, the replacement of $\eta$ by $\eta_{\mathrm{eff}}$ in \autoref{eq: ohm_full} can be regarded as the definition of an effective resistivity that incorporates kinetic effects in single-fluid resistive MHD. As we justify in \autoref{app:full_Ohm}, we can further assume that $|\rho_e {\bf v}|\ll |{\bf J}|$ and $\Gamma\simeq 1$, and that the third term in the square bracket is negligible, which yields
\begin{equation}\label{Eq:eta}
    \mathbf{E}^*\equiv \mathbf{E}+\frac{\mathbf{v}}{c} \times \mathbf{B} = \eta_{\mathrm{eff}} \mathbf{J}
\end{equation}
where $\mathbf{E}^*$ is the nonideal electric field. The spatial structure of the $z$-component of the nonideal electric field, $E_z^*$, is shown in the bottom row of \autoref{fig:sim_setul}, at a representative time after the simulation has achieved a quasi-steady state {(i.e., the reconnection rate attains a quasi-steady value)}. The figure emphasizes that nonideal regions are generally larger for increasing guide field. We also present the spatial structure of the total particle density $n_t$ (top row; in units of $n_0$) and of the magnetic energy density (middle row; in units of $B_0^2/8\pi$), for both $B_g/B_0=0$ (left column) and $B_g/B_0=1$ (right column). 

In order to determine the effective resistivity $\etaeff$, we focus on the $z$-component of \autoref{Eq:eta}, which dominates the nonideal field for the whole  range of $B_g/B_0$ we explore. The $z$-component $E_z^*$ is the only significant component for zero guide field, {being one to two orders of magnitude larger than other components}; for non-zero guide fields, we still determine $\etaeff$ from $E_z^*=\etaeff J_z$, but we show that the same effective resistivity properly describes other components, specifically $E_x^*=\etaeff J_x$ (see \autoref{fig:x_comp}). Our new prescription for the effective resistivity is derived using a data-driven phenomenological model with two free parameters, which are benchmarked with PIC simulations. We compare the performance of our prescription to the resistivity model from \cite{Selvi}---based on a kinetic approach---and to its extension employing MHD quantities.

\begin{figure}
    \centering
    \includegraphics[width=0.9\linewidth]{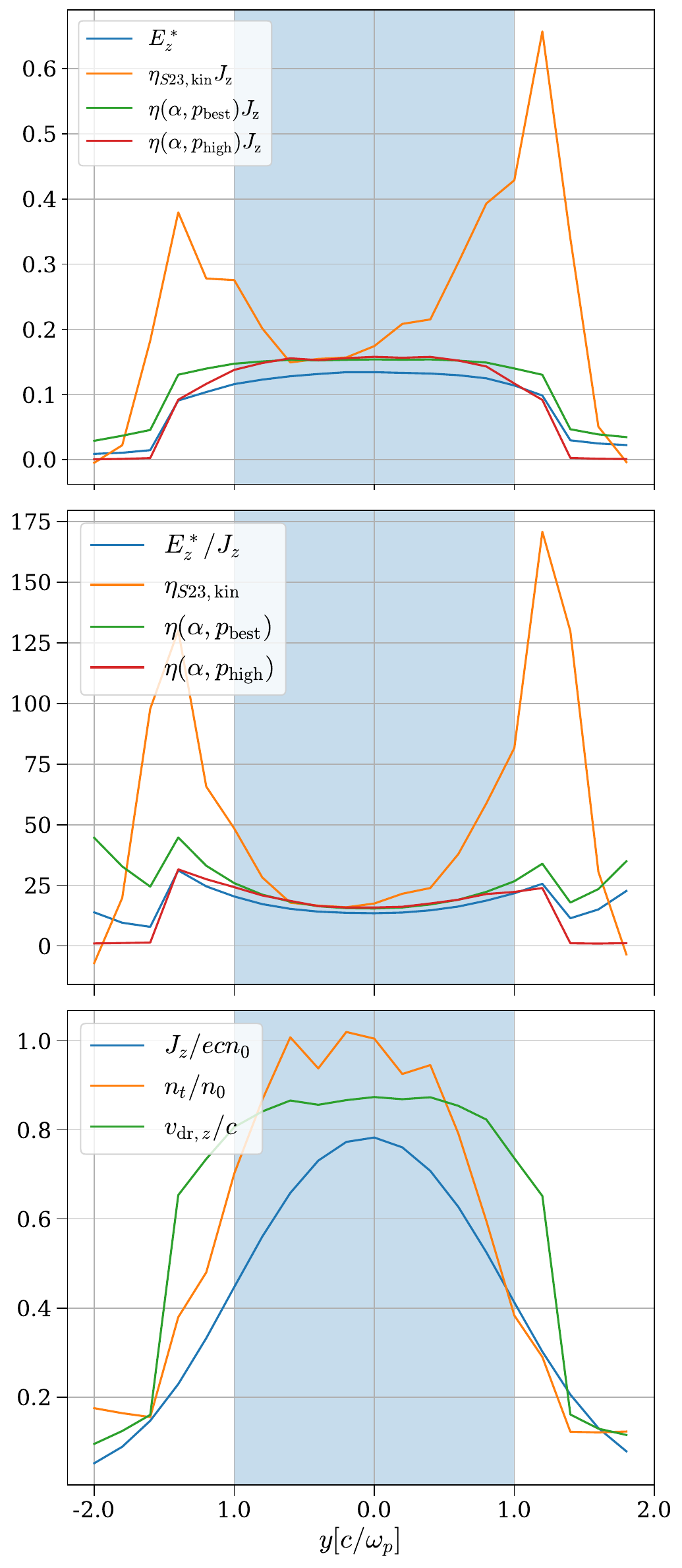}
    \caption{1D slice of the domain along $y$ through an X-point, for the simulation with zero guide field. The top panel shows the {$z$ component of the} nonideal {electric} field {in units of $B_0$}, the second panel the resistivity, and the bottom panel the electric current $J_z$ (in blue), the number density $n_t$ (in orange) and the drift speed $v_{{dr},z}/c\simeq J_z/en_t c$ (in green). In the top and middle panels, we present in blue the ground truth obtained directly from our simulation, while other colors show various choices for $\etaeff$, as described in the legend. {Our prescription for resistivity (\autoref{Eq:Guess}) is shown as $\eta(\alpha, p_{\text{best}})$ and $\eta(\alpha, p_{\text{high}})$ for the values of $p$ defined in \autoref{sec:results} and corresponding $\alpha$ values.}
    Regions where $|\mathbf{E}|>|\mathbf{B}|$ are shaded in blue.}
    \label{fig:1D}
\end{figure}

\begin{figure*}
    \centering
    \includegraphics[width=.8\linewidth]{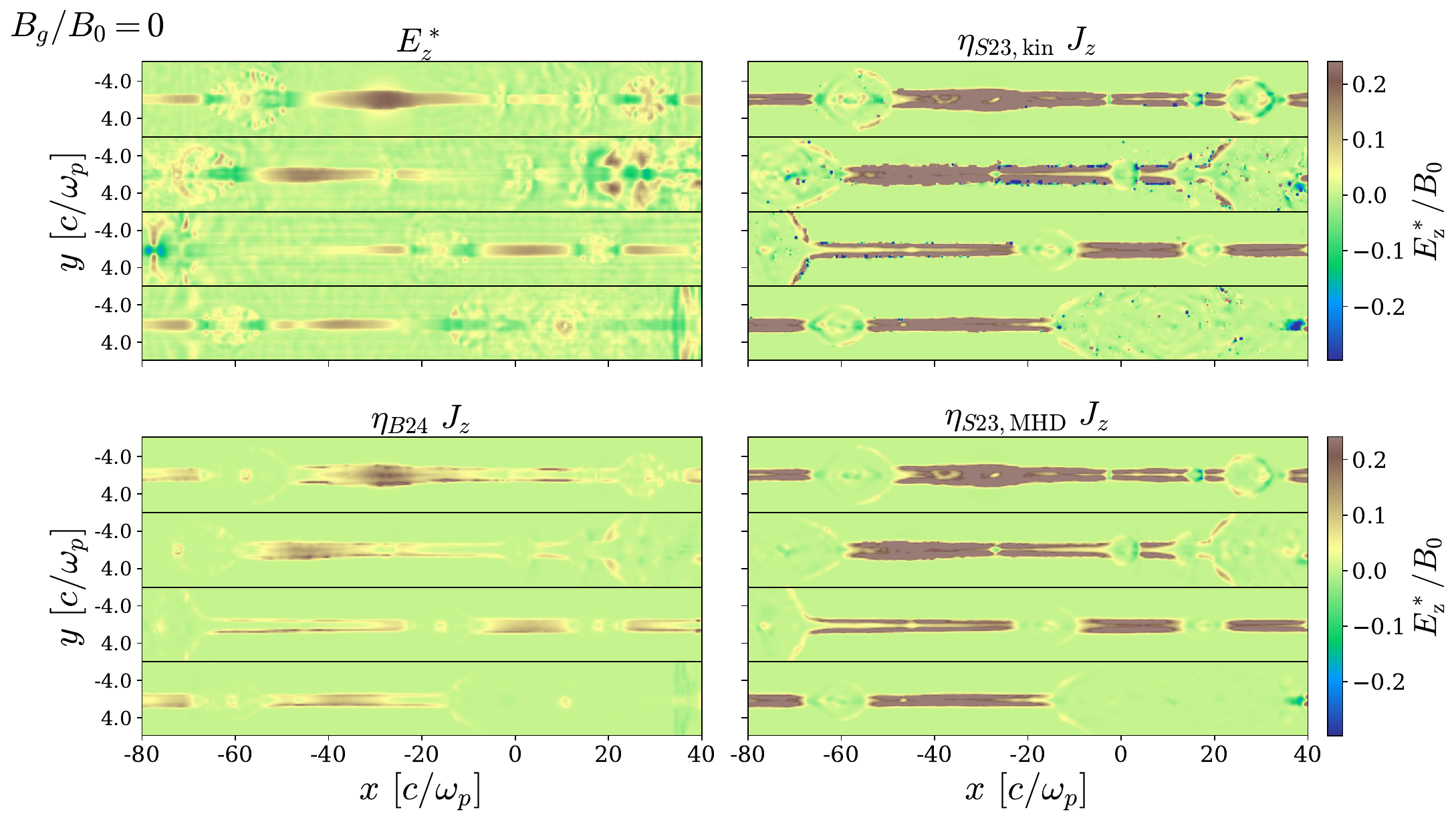}
    \caption{A comparison between the measured nonideal electric field $E_z^*$ (top left) and its reconstruction $\etaeff J_z$ based on different choices of $\etaeff$: $\eta_{\text{S23,kin}}$ (\autoref{Eq:Selvi}) in top right, $\eta_{\text{B24}}$ (\autoref{Eq:bugli}) in bottom left, and $\eta_{\text{S23,MHD}}$ (\autoref{Eq:SelviMHD}) in bottom right. All panels are normalized to $B_0$. Within each panel, horizontal black lines separate different time snapshots: the first one is taken when the reconnection rate shown in \autoref{fig:reconnection} first settles into a steady state, and the others follow after 450, 810, and 1080 $\omega_p^{-1}$ respectively. The horizontal axis is measured with respect to the center of the portion of  domain that is displayed. The derivatives in \autoref{Eq:Selvi}, \autoref{Eq:SelviMHD} and \autoref{Eq:bugli} are computed as numerical derivatives on cells downsampled by a factor of two.
   }
    \label{fig:selvi_comp1}
\end{figure*}

\subsection{Kinetically motivated resistivity}\label{subsec:selvi_method}
\cite{Selvi} analyzed PIC simulations of relativistic reconnection in pair plasmas and identified the terms that dominate the nonideal electric field in the generalized Ohm's law \citep{Hesse07}. Their analysis was restricted to regions of electric dominance, defined as having $E_z^2> B_x^2+B_y^2$ (which is nearly identical to the condition $|\mathbf{E}| > |\mathbf{B}|$ in the case of zero guide field). They found that the $z$-component of the nonideal electric field could be written as
    \begin{equation} \label{Eq:Selvi}
        E_z^* = \eta_{S23,\mathrm{kin}}J_z = \left[\frac{m}{n_t e^2}\frac{\langle u_{e z}\rangle}{\langle v_{e z}\rangle}\partial_y\langle v_{ey} \rangle\right]J_z 
    \end{equation}
where $n_t$ is the total number density (including both electrons and positrons), $\langle v_{e z}\rangle$  and $\langle u_{e z}\rangle$ are respectively the mean electron three- and four-velocity in the $z$-direction,\footnote{At X-points, positrons and electrons have opposite $\langle v_{e z}\rangle$  and $\langle u_{e z}\rangle$, but the ratio $\langle u_{e z}\rangle/\langle v_{e z}\rangle$ is the same for both species.} and $\langle v_{ey} \rangle\simeq v_{y} $ is the mean three-velocity along $y$, which is roughly the same for both species (hereafter, we call $v_{y}$ the single-fluid $y$ velocity).

The effective resistivity proposed by \cite{Selvi} in \autoref{Eq:Selvi} has a few limitations: (\textit{i}) it provides a satisfactory description of the nonideal electric field only for $B_g=0$; and (\textit{ii}) it was derived considering  regions of electric dominance, which are only a subset of the regions hosting nonideal fields \citep{Sironi2022,totorica_23}, where resistive effects are important. In order to derive \autoref{Eq:Selvi}, \cite{Selvi} used the approximation
\begin{equation}\label{Eq:approx} \partial_y \left(n_e \langle v_{ey} \rangle \langle u_{ez} \rangle \right) \approx n_e \langle u_{ez} \rangle \partial_y \langle v_{ey}\rangle
\end{equation} 
which is valid only in the vicinity of {the center of the current sheet}. 
In fact, as shown in \autoref{fig:1D}, the effective resistivity in \autoref{Eq:Selvi}  (hereafter, $\eta_{S23,\rm kin}$) provides a reasonable description of the nonideal  field near the center of the layer ($|y|\omega_p/c\lesssim 1$), where $|\mathbf{E}| > |\mathbf{B}|$  (blue shaded area), but it significantly overestimates the ground truth ({i.e., the direct measurement of $E_z^*$ from PIC runs}) farther away from the layer ($|y|\omega_p/c\gtrsim 1$). 

For use in single-fluid MHD codes, \autoref{Eq:Selvi} needs to be rewritten using  fluid quantities. As we have already discussed above, the mean three-velocity along $y$ is roughly the same for the two species, $\langle v_{ey} \rangle\simeq\langle v_{y} \rangle$. The most reasonable approximation for the ratio between the mean four- and three-velocities of a given species is $\langle u_{ez} \rangle/\langle v_{ez} \rangle\simeq \langle \gamma \rangle$, where the mean particle Lorentz factor (including both bulk and internal motions) can be derived from the $T^{00}$ component of the stress energy tensor as $\langle \gamma \rangle=T^{00}/n_t mc^2$. This leads to a 
form of \autoref{Eq:Selvi} that can be implemented in MHD:
\begin{equation} \label{Eq:SelviMHD}
    E_z^*=\eta_{S23, \rm MHD}J_z =\left[\frac{m}{n_t e^2}\langle \gamma\rangle \ \partial_y v_y \right] J_z~. 
\end{equation}
As shown in \autoref{fig:selvi_comp1}, \autoref{Eq:SelviMHD} is an excellent approximation of the kinetic form in \autoref{Eq:Selvi} (compare top right and bottom right panels). However, as anticipated in \autoref{fig:1D}, the two forms overestimate the true resistivity (top left of \autoref{fig:selvi_comp1}), especially at the boundaries of the current layer. In an earlier version of \citet{Selvi}, \autoref{Eq:Selvi} was cast in an alternative form, approximating
\begin{equation}
\label{eq:approx}
    \frac{\langle u_{ez} \rangle}{\langle v_{ez} \rangle}\simeq \frac{1}{\sqrt{1-(J_z/en_t c)^2}}~,
\end{equation}
which only holds if each species has negligible internal motions and moves in the $z$ direction with dimensionless drift speed of $|J_z|/e n_t c$. This was recently rewritten by \citet{Bugli} in the form 
\begin{equation}\label{Eq:bugli}
    \eta_{B24} = \frac{1}{en_t c}\sqrt{\left(\frac{mc}{e}\partial_y v_y\right)^2+\left(\Gamma E^*_z\right)^2}~.
\end{equation}
While the approximation in \autoref{eq:approx} leading to \autoref{Eq:bugli} does not generally hold, as shown by the poor agreement between the top right  and  bottom left panels in \autoref{fig:selvi_comp1}, \autoref{Eq:bugli} appears to provide a remarkably good proxy for the ground truth (compare top left and bottom left). While \autoref{Eq:bugli} appears to improve upon the kinetically-motivated model by \citet{Selvi}, it loses some of the physical motivation of \autoref{Eq:Selvi} and \autoref{Eq:SelviMHD}.

While useful, the forms of effective resistivity presented in this subsection have some undesirable properties: (\textit{i}) they only apply to the case of zero guide field; (\textit{ii}) they contain a spatial derivative, which makes them difficult to include in relativistic MHD codes while maintaining causality \citep{DelZanna}; (\textit{iii}) they only apply to the main layer, and not to the anti-reconnection layers in between merging plasmoids (which extend along $y$, and for which the relevant velocity derivative is $\partial_x v_x$);
(\textit{iv}) they retain a dependence on the system geometry (e.g., via the $z$ component $E_z^*$), which makes it hard to incorporate in global MHD simulations where current sheets will be curved, oscillating, and generally not aligned with the coordinate axes. In the next subsection we turn to a more agnostic approach that avoids some of these issues.

\begin{figure}
    \centering
    \includegraphics[width=1\linewidth]{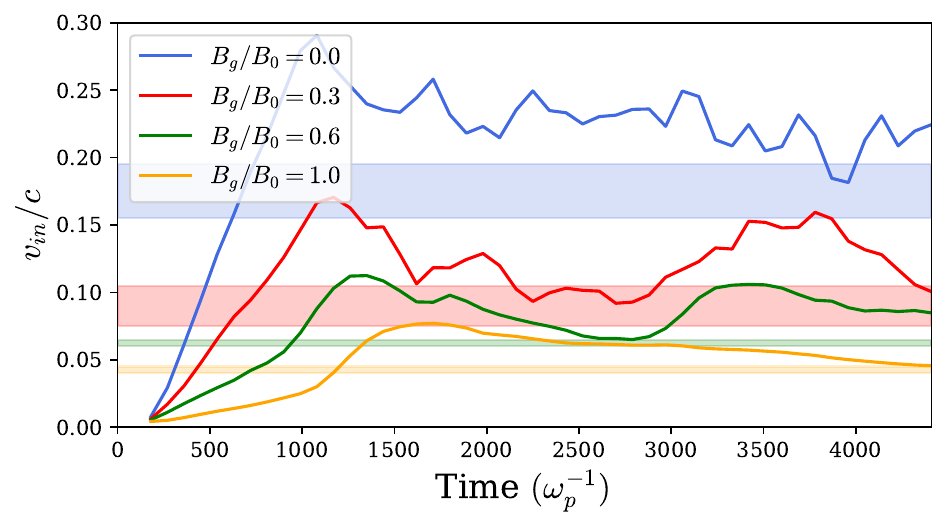}
    \caption{Reconnection rate (i.e., the plasma inflow velocity normalized by $c$) over time for each guide field case, as indicated in the legend. The reconnection rate is measured as the mean inflow velocity in the region $y=[-672, 672] c/\omega_p$.
    In the same color we show with the shaded area the acceptable ranges of $\alpha/2$ of our resistivity prescription, which we define in \autoref{subsec:guess_method}.}
    \label{fig:reconnection}
\end{figure}

\subsection{Prescriptive resistivity}\label{subsec:guess_method}
To overcome the limitations of the model by \citet{Selvi}, we propose an empirical approach. We expect that in regions of strong current---as defined below---the nonideal electric field should approach $|{\bf E^*}|\rightarrow (v_{in}/c)B_0$, where $v_{in}$ is the reconnection rate (i.e., the inflow velocity of plasma into the layer, see \autoref{fig:reconnection}), which implies that the effective resistivity should be
\begin{equation}
\etaeff\rightarrow \frac{v_{in}}{c} \frac{B_0}{|\mathbf{J}|}    ~.
\end{equation}
A choice of $\etaeff\propto |\mathbf{J}|^{-1}$ is inapplicable in regions of small electric current, where the resistivity should vanish. We therefore design a form such that $\etaeff\propto |\mathbf{J}|^p$ for small current densities, where $p>0$ is a free parameter. More precisely, this should occur where $|\mathbf{J}|\ll e n_t c$. Adding a normalization factor $\alpha$, this motivates choosing a form

\begin{equation} \label{Eq:first_guess}
\eta_{\mathrm{eff}} =  \frac{ \alpha B_0 \mathbf{|J|}^p}{|\mathbf{J}|^{p+1}+\left(e n_t c \right)^{p+1}}=\frac{ \alpha B_0}{|\mathbf{J}|\left[1+\left({en_tc}/{|\mathbf{J}|}\right)^{p+1}\right]}.
\end{equation}
We will determine free parameters $\alpha$ and $p$ from PIC simulations. This scales as $\etaeff\propto |\mathbf{J}|^p /(e n_t c)^{p+1}$ at small currents and approaches
 $\eta_{\mathrm{eff}}=\alpha B_0/(2 |\mathbf{J}|)$ for $|\mathbf{J}|\simeq e n_t c$. We therefore expect $\alpha/2\simeq v_{in}/c$, as we indeed find below (see also \autoref{fig:reconnection}). 
 The condition $|\mathbf{J}|\simeq e n_t c$ corresponds to the charge starvation regime, i.e., all charge carriers move at near the speed of light. This limit is indeed realized in the inner region of the current sheet: as the bottom panel of \autoref{fig:1D} shows, the 1D profiles of $J_z$ and $n_t$ have the same shape, suggesting a nearly constant drift velocity $J_z/e n_t\simeq 0.9\,c$. In fact, if we define the drift velocity ${\bf v}_{dr}\equiv {\bf J}/e n_t$, our prescription can be written as
\begin{equation} \label{Eq:Guess}
\eta_{\mathrm{eff}} = \frac{ \alpha B_0}{|\mathbf{J}|\left[1+\left({c}/{|{\bf v}_{dr}|}\right)^{p+1}\right]}~.
\end{equation}
In the inner region of the current sheet, where $|{\bf v}_{dr}|\simeq c$ (green line in the bottom panel of \autoref{fig:1D}), we obtain  $\etaeff\propto |{\bf J}|^{-1}$, which matches the double-peaked shape of the ground truth (i.e., $E_z/J_z$) in the middle panel of \autoref{fig:1D}. 
We emphasize that the density dependence in ${\bf v}_{dr}\propto {\bf J}/n_t$ is a key ingredient of our resistivity model---in fact, the density in the middle of the sheet  can be significantly larger than in the immediate upstream, see bottom panel of \autoref{fig:1D}.

In Section \ref{sec:discussion} we provide an equivalent, more general version of Equation~\eqref{Eq:first_guess} suitable for implementation within resistive MHD codes.

\begin{figure}
    \centering
    \includegraphics[width=\linewidth]{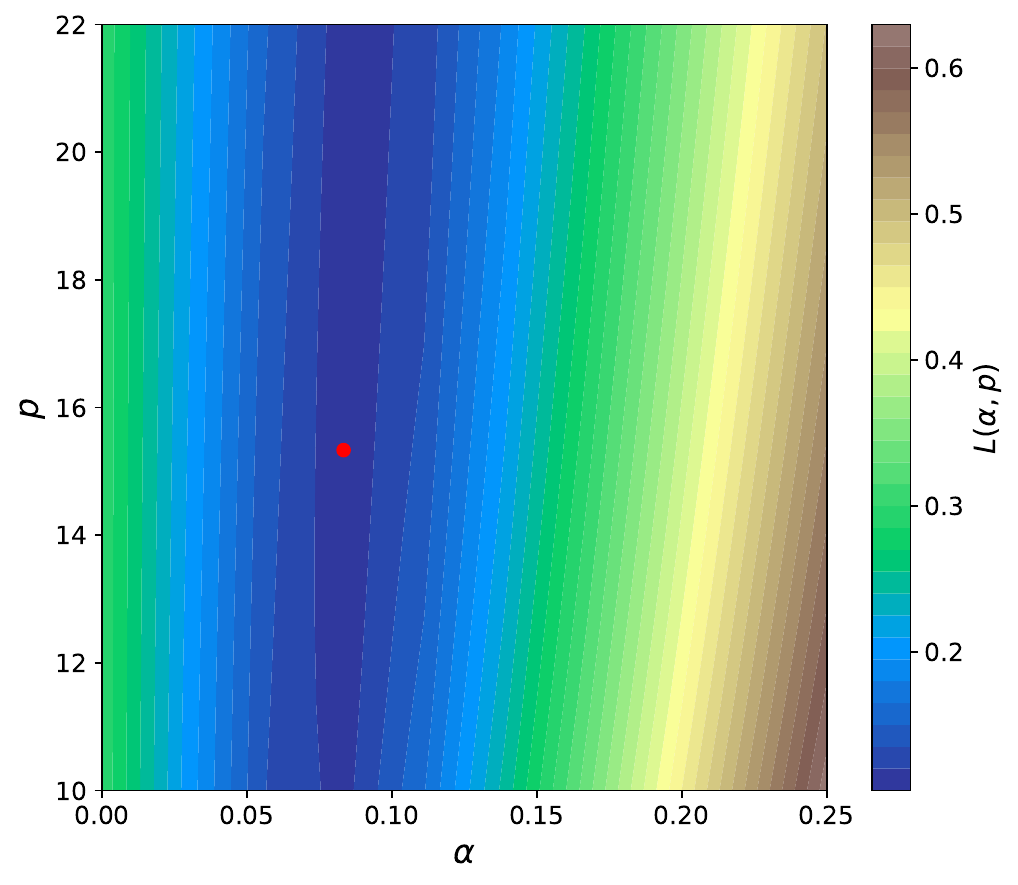}
    \caption{Loss manifold computed with \autoref{eq:loss} for $B_g/B_0=1$, as a function of $\alpha$ and $p$. The red point marks the $(\alpha, p)$ combination of minimum loss. Many combinations in the dark blue valley produce losses very close to the global minimum. A correlation between the two parameters can also be seen.}
    \label{fig:loss_curve}
\end{figure}

\begin{table}[]
    \centering
    \renewcommand{\arraystretch}{1.4} 
    \begin{tabular}{|c|c|c|c|}
    \hline
    \hline
    $B_g/B_0$  &Patch Dim. $(c/\omega_p)$ & Num. Patches & $p$\\
    \hline
        $0.0$ & $[80, 16]$& $2500$& $0.00^{+1.73}_{-0.00}$  \\
        $0.3$ & $[160, 40]$ & $2000$& $9.59^{+8.59}_{-5.37}$\\
        $0.6$ & $[240, 40]$ & $600$& $15.4^{+3.8}_{-2.1}$\\
        $1.0$ & $[120, 24]$ & $1500$& $18.2^{+5.1}_{-6.2}$\\
        \hline
        \hline
    \end{tabular}
    \caption{Best fit $p$ with {upper and lower limits of the acceptable range} (last column) for each guide field. We indicate the dimensions (along $x$ and $y$ respectively) of the patches used to compute the distribution of values of $p$ (second column) as well as the number of patches (third column). }
    \label{tab:search_params}
\end{table}

\begin{table}
    \centering
    \renewcommand{\arraystretch}{1.3} 
    \begin{tabular}{|c|c|c|c|}
    \hline
    \hline
        $B_g/B_0$ & $p_{\mathrm{best}}$& $[p_{\mathrm{low}}, p_{\mathrm{high}}]$  & $\alpha (p)$ \\
        \hline
        \hline
        $0.0$ &$0.00$ &$[0.00, 1.73]$ & $0.0369\,p + 0.3268$\\
        $0.3$ &$9.59$& $[4.22, 18.2]$ & $0.0046\, p + 0.1295$\\
        $0.6$ &$15.4$& $[13.3, 19.2]$ & $0.0017\, p + 0.1002$\\
        $1.0$ &$18.2$& $[12.0, 23.3]$ & $0.0010 \,p + 0.0702$\\
        \hline
        \hline
    \end{tabular}
    \caption{For each guide field we show the best fit value of $p$ (second column), and the lowest and highest acceptable values (third column). These bounds are calculated as described in the text. We also show a function which returns the optimal $\alpha$ for a given $p$ within this range (fourth column). }
    \label{tab:alpha_of_p}
\end{table}

\begin{table}[]
    \centering
    \renewcommand{\arraystretch}{1.4} 
    \begin{tabular}{|c|c|c|c|}
    \hline
    \hline
    $B_g/B_0$ &Patch Dim. $(c/\omega_p)$ & Threshold per. & $p$ \\
    \hline
        $0.0$  &$[60, 32]$& $55-45$ & $0.00^{+1.26}_{-0.00}$ \\
        $0.0$  &$[80, 16]$&$60-40$ & $0.00^{+1.33}_{-0.00}$ \\
        $1.0$  &$[80, 40]$& $55-45$ & $17.8^{+5.4}_{-5.9}$ \\
        $1.0$  &$[120, 24]$&$60-40$ &  $18.2^{+4.8}_{-4.8}$ \\
        \hline
        \hline
    \end{tabular}
    \caption{The results obtained when varying the size of patches used to compute $p$ (default values are in \autoref{tab:search_params}) and the percentiles used in the threshold for patch selection (default values are $55-45$). As before, for $B_g/B_0 =0.0$ we use 2500 patches and for $B_g/B_0 =1.0$, 1500 patches.}
    \label{tab:tests}
\end{table}

\begin{figure}
    \centering
    \includegraphics[width=.99\linewidth]{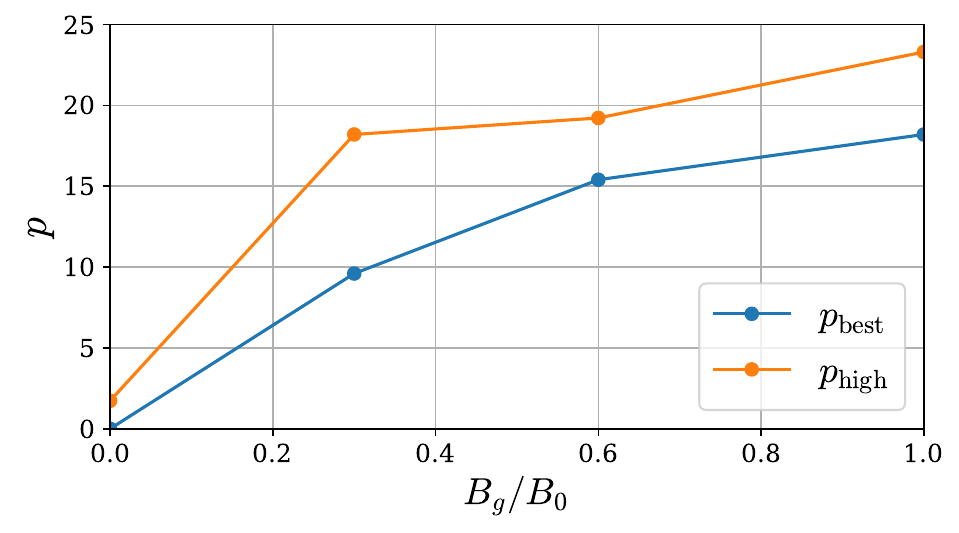}
    \caption{Best fit ($p_{\rm best}$) and upper bound ($p_{\rm high}$) as a function of guide field strength. }
    \label{fig:p(pg)}
\end{figure}

\begin{table}
\centering
\begin{center}
\begin{tabular}{|c|c|c|c|c|}
    \hline
    $B_g/B_0$ & Model & $|E_z^*|$  & $(E_z^*)^2$  & 1\\
    \hline
    \hline
   \renewcommand{\arraystretch}{1.8} 
    \multirow{2}{*}{$0.0$}
        &$\eta (\alpha, p_{\text{best}}) J_z$ & 5.052  &3.517 & 21.35\\
        &$\eta (\alpha, p_{\text{high}}) J_z$ &  5.157 &3.599 & 17.93\\
    \hline
    \multirow{2}{*}{$0.3$}
    \renewcommand{\arraystretch}{1.5} 
        &$\eta (\alpha, p_{\text{best}}) J_z$ & 3.121  &1.085 & 12.97\\
        &$\eta (\alpha, p_{\text{high}}) J_z$ &  3.129 &1.086 & 13.00\\
    \hline
    \multirow{2}{*}{$0.6$}
        &$\eta (\alpha, p_{\text{best}}) J_z$ & 0.410  &0.067 & 3.913\\
        &$\eta (\alpha, p_{\text{high}}) J_z$ &  0.412 &0.067 & 3.920\\
    \hline
    \multirow{2}{*}{$1.0$}
        &$\eta (\alpha, p_{\text{best}}) J_z$& 0.108  &0.008 & 2.047\\
        &$\eta (\alpha, p_{\text{high}}) J_z$ &  0.110 &0.009 & 2.048\\

    \hline
    \hline
\end{tabular}
\end{center}
\caption{Performance L2 loss of various resistivity models as compared to the measured $E_z^*$, for different guide fields. We vary the weight of the loss function as indicated in the last three columns. We exclude cells where $n_t<1$.}
\label{tab:loss}
\end{table}

\section{Results} \label{sec:results}
 To determine the optimal values of $\alpha$ and $p$  in \autoref{Eq:Guess} we consider the $z$ component of the nonideal field and
 define a loss, or data-fit metric
\begin{equation}\label{eq:loss}
    L(\alpha, p) = \sum_{x,y} |E_z^*-\etaeff(\alpha, p)  J_z|^2|E_z^*|
\end{equation}
i.e., we minimize the L2 loss (the mean squared error) between $\eta_{\mathrm{eff}} J_z $ and the measured $E_z^*$.
The loss is weighted by $|E_z^*|$ to ensure that the large regions with negligible nonideal fields do not skew our findings. We calculate the optimal parameters $\alpha$ and $ p$  by minimizing this loss through a simple grid search. We create a composite domain including several time snapshots of the PIC simulations. {For each case with varying guide field}, the snapshots (roughly 15 in each case) are equally spaced from the time when reconnection first attains a quasi-steady state up to the end of our simulations, $\omega_p t=4500$. 
For each snapshot, we consider a region extending along the whole domain in $x$ and with thickness $64\,c/ \omega_p$  along $y$ (sufficient {to enclose the largest plasmoids}), centered around the current sheet. As a representative case, the loss for $B_g/B_0=1$ is shown in \autoref{fig:loss_curve}. The manifold shows that a valley of small loss, with values of $L$ near the global minimum (shown by the red point), stretches across a wide range of $p$. 

In order to determine the optimal $p$ and {define a range of acceptable values} we adopt the following procedure. We begin by minimizing the loss on many small, randomly selected regions (hereafter, ``patches'') of the composite domain. {In each small patch we find that there is a clearly preferred value of $p$ (i.e. a sharp minimum of \autoref{eq:loss}, as opposed to the wide minimum we find on global scales) which we will use to define a range of acceptable values of $p$.}
We continue adding regions until the results converge, meaning that repeatedly selecting the same number of random patches produces the same outcome, regardless of which regions are chosen.
We vary the patch size depending on the guide field strength, such that the patch is twice larger than the typical extent of a region with significant nonideal fields (see bottom panels in  \autoref{fig:sim_setul}). The number of patches and the patch size used in this step are indicated in \autoref{tab:search_params}. To ensure that the loss in a given patch is informative (which is not the case for patches with small $E_z^*$), we require 
\begin{equation}  
\label{eq:patch}
\mathrm{median}(E_{z}^*)_{\rm patch} > \mathrm{median}(E^*_{z})_{\rm global}+ \mathrm{threshold}  
\end{equation}
where the median is computed in a given patch (left hand side) or over the whole composite domain (right hand side). The threshold indicated on the right hand side is the difference between the 55th and 45th percentiles of the distribution of $E_{z}^*$ in the whole domain. We use these percentiles instead of the standard interquartile range to ensure a more robust analysis that includes a greater portion of the domain. We define $p_{\rm best}$ as the value that minimizes the loss when considering the combined area of all patches.

\begin{figure*}
    \centering
    \includegraphics[width=\linewidth]{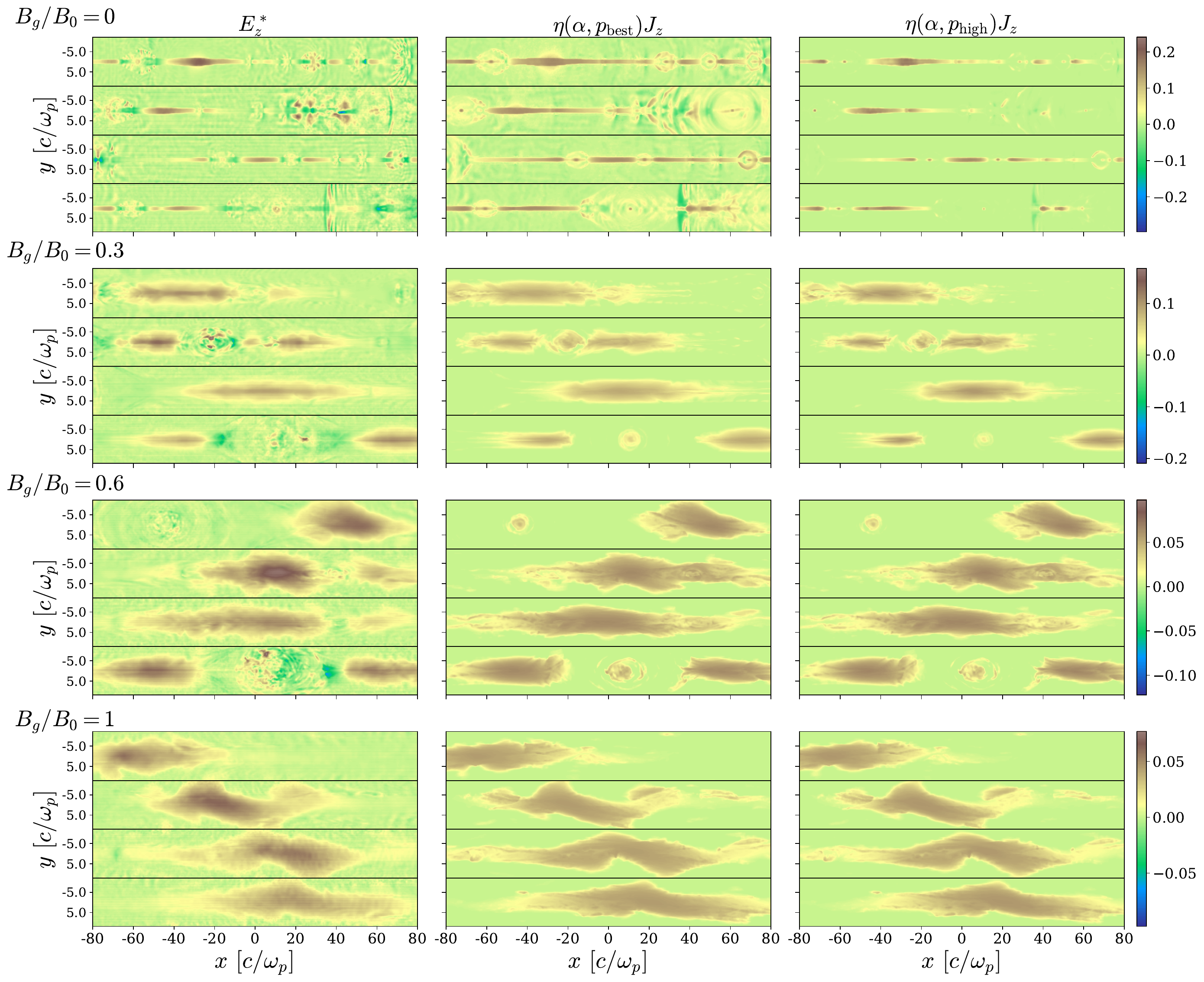}
    \caption{A comparison between the measured nonideal electric field $E_z^*$ (left column) and its reconstruction $\etaeff J_z$ based on our prescription in \autoref{Eq:Guess}, for the whole range of guide fields we explored. The middle  column shows $\etaeff(\alpha,p_{\rm best})J_z$ (here, $\alpha$ is the value corresponding to $p_{\rm best}$ based on the linear fit in \autoref{tab:alpha_of_p}), while the right column shows $\etaeff(\alpha,p_{\rm high})J_z$ ($\alpha$ is the value corresponding to $p_{\rm high}$).
    All panels are normalized to $B_0$. Within each panel, horizontal black lines separate different time snapshots: the first one is taken when the reconnection rate shown in \autoref{fig:reconnection} first settles into a steady state, and the others follow after 450, 810, and 1080 $\omega_p^{-1}$ respectively. The horizontal axis is measured with respect to the center of the portion of  domain that is displayed (which is a small fraction of the composite domain used to determine the best-fit values of $\alpha$ and $p$). 
}
    \label{fig:guess_comp}
\end{figure*}

\begin{figure*}
    \centering
    \includegraphics[width=\linewidth]{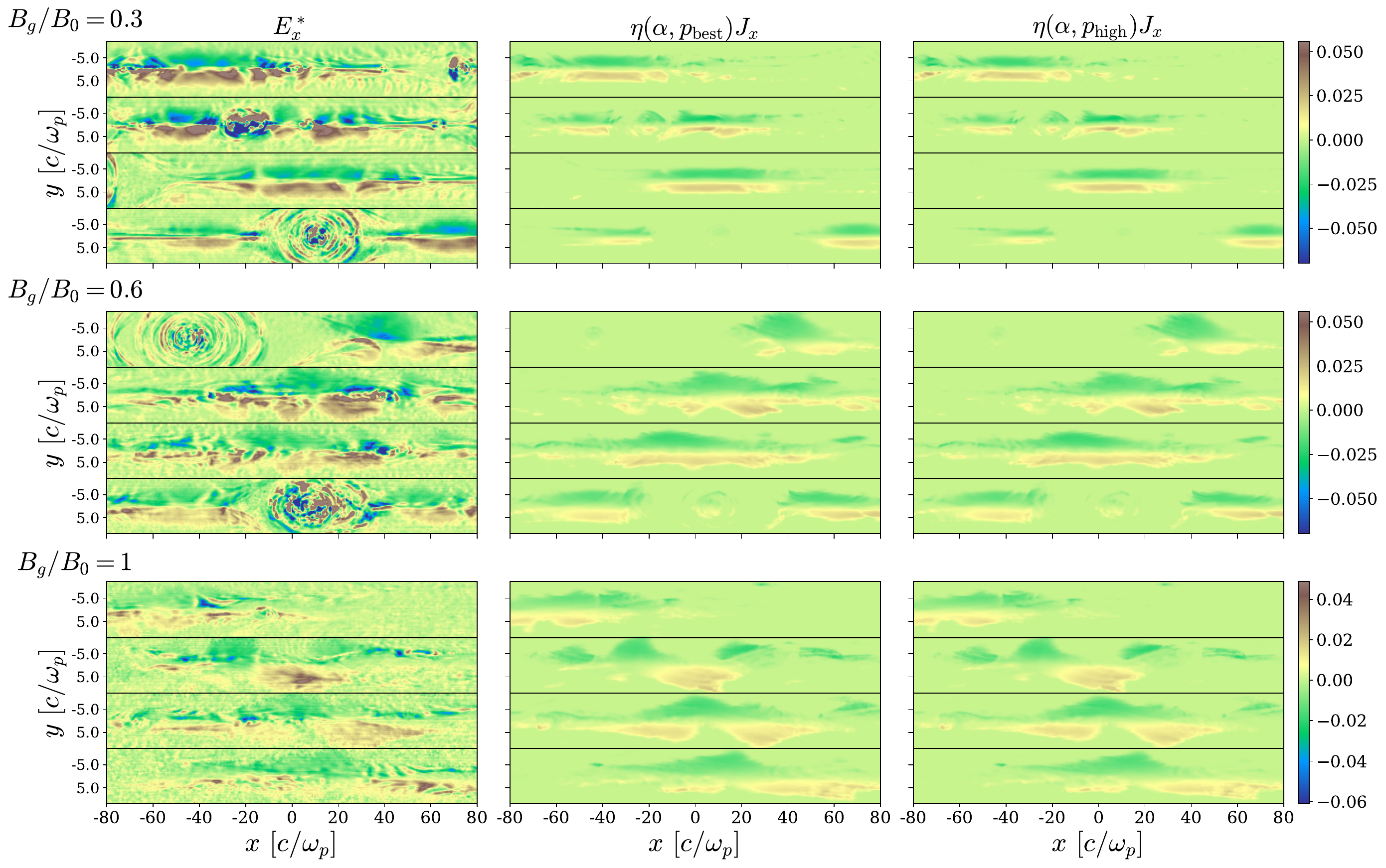}
    \caption{Same as \autoref{fig:guess_comp}, but for the $x$ component of the nonideal field, which appears for non-zero guide field cases. The resistivity is based on \autoref{Eq:Guess} and employs the same values of $\alpha$ and $p$ as in  \autoref{fig:guess_comp}. Although our model was not developed using
    the $x$ component of $\mathbf{E}^*$, it is able to recover the basic structure of regions with significant $E_x^*$. The magnitude of $E_x^*$ is in units of $B_0$, and the times shown are the same as in \autoref{fig:guess_comp}. 
    }
    \label{fig:x_comp}
\end{figure*}

We then create the distribution of values of $p$ that minimize the loss in each patch. 
The difference $\Delta p_{\rm low}$ between the 50th and 16th percentiles of this distribution gives the lower limit on allowed values of $p$, $p_{\text{low}}=\max[p_{\rm best}-\Delta p_{\rm low},0]$, while the difference $\Delta p_{\rm high}$ between the 84th and 50th percentiles gives the upper limit $p_{\text{high}}=p_{\rm best}+\Delta p_{\rm high}$. These values are indicated for each guide field in \autoref{tab:search_params}. We plot $p_{\mathrm{best}}$ and $p_{\mathrm{high}}$ as a function of guide field in \autoref{fig:p(pg)}. \autoref{tab:tests} demonstrates that our findings do not depend much on the patch size or the threshold percentiles employed in \autoref{eq:patch}. We find that the optimal value of $p$ is robust to patch size, and varies by $\lesssim 2\%$ when the threshold percentiles are altered. Similarly, the {upper and lower limits on} on $p$ decrease by a modest amount when varying these parameters.

Within the range $[p_{\rm low},p_{\rm high}]$, we then consider 15 evenly spaced values of $p$. For each value, we find the optimal $\alpha$ using the loss function on the combined area of all patches. This reveals that the two parameters are correlated. We interpolate to find $\alpha(p)$ and show the resulting linear fits in \autoref{tab:alpha_of_p}. The range of $\alpha/2$ corresponding to the interval $[p_{\rm low},p_{\rm high}]$ is shown by the colored shaded bands in \autoref{fig:reconnection}. As expected, $\alpha/2$ matches  well with the measured reconnection rate, for all guide field cases (solid lines of the same color).

We finally assess how well the reconstructed $\etaeff J_z$ captures the nonideal field $E_z^*$ obtained directly from our PIC simulations. \autoref{tab:loss} shows the L2 loss obtained for $p=p_{\rm best}$ or $p=p_{\rm high}$. For each of the two choices, the corresponding value of $\alpha$ is obtained from the linear fit in \autoref{tab:alpha_of_p}. We find that, regardless of the weight adopted in the loss function (no weight, $|E_z^*|$ or $(E_z^*)^2$), the L2 loss increases by less than $\sim 10\%$ when using $p_{\rm high}$, as compared to choosing $p_{\rm best}$ (and for the unweighted loss of $B_g/B_0=0$, $p_{\rm high}$ performs better than $p_{\rm best}$). We therefore regard all solutions within the range of $[p_{\rm best}, p_{\rm high}]$ as acceptable.

This is also confirmed by the 2D spatial profiles shown in \autoref{fig:guess_comp}. For all the guide fields we explore, we present the ground truth in the left column (i.e., the nonideal field $E_z^*$ measured directly from our simulations), the reconstruction $\etaeff(\alpha,p_{\rm best})J_z$ in the middle column (here, $\alpha$ is the value corresponding to $p_{\rm best}$ based on the linear fit in \autoref{tab:alpha_of_p}), and the  reconstruction $\etaeff(\alpha,p_{\rm high})J_z$ in the right column (here, $\alpha$ is the value corresponding to $p_{\rm high}$). The plot shows that the two reconstructions are equally good for non-zero guide fields, while for $B_g/B_0=0$ the case $\etaeff(\alpha,p_{\rm high})J_z$ seems to capture better the longitudinal extent of nonideal regions. Most importantly, our prescriptive resistivity performs clearly better than the kinetically-motivated models presented in \autoref{fig:selvi_comp1}.

Although our prescriptive resistivity was benchmarked with the $z$ component of the nonideal  field, it can successfully model other non-trivial components that appear for non-zero guide fields. This is shown in \autoref{fig:x_comp}.
While $E_y^*$ is negligible for all guide fields, there are distinct areas in which $E_x^*$ is significant for  non-zero guide field cases. We calculate $\eta_{\text{eff}}J_x$ via \autoref{Eq:Guess} using the same $\alpha$ and $p$ from the analysis of the $z$-component described above. From the results in \autoref{fig:x_comp} we can conclude that the scalar resistivity in \autoref{Eq:Guess} provides a satisfactory description of all components of the nonideal electric field, across the whole range of guide fields that we explore.

\section{Discussion}\label{sec:discussion}
We have performed a suite of PIC
simulations of relativistic pair-plasma reconnection with varying guide field strength, and we have formulated an empirical prescription for the effective resistivity $\etaeff$ in \autoref{Eq:first_guess} or equivalently \autoref{Eq:Guess}. Our prescription depends on two free parameters, $\alpha$ and $p$, which are derived directly from our PIC runs ---with $\alpha/2$ expected to be comparable to the dimensionless reconnection rate.
 As compared to the kinetically-motivated model proposed by \citet{Selvi}, the form of $\etaeff$ that we propose has four main advantages: it is explicitly written in single-fluid MHD quantities, does not depend on spatial derivatives, is coordinate-agnostic, and is valid for any guide field. It depends only on the electric current density and the particle number density (and the two free parameters discussed above). We have demonstrated that the scalar resistivity we propose successfully describes the spatial structure and strength of all components of the nonideal field. It thus provides a
promising strategy for enhancing the reconnection rate in relativistic resistive MHD approaches.

To confirm the robustness of our findings, we demonstrate in \autoref{app:tests} that the form in \autoref{Eq:Guess} (with $\alpha$ and $p$ determined from our reference runs) provides an excellent description of nonideal fields in independent simulations which either include synchrotron cooling or resolve the plasma skin depth with $20$ cells (as compared to $5$ cells for our reference runs).

We conclude with an important remark. Our prescription in \autoref{Eq:first_guess} can be equivalently written as 
\begin{equation}\label{eq:eta_eff_alt}
    \eta_{\text{eff}} = \frac{|\mathbf{E}^*|}{e n_t  c}\left[\frac{\alpha B_0 - |\mathbf{E}^*|}{|\mathbf{E}^*|} \right]^{\frac{1}{p+1}}.
\end{equation}
In the limit of very high $p$, the square bracket is elevated to a very small power, yielding a contribution of order unity. Furthermore, \autoref{fig:loss_curve} suggests that, as long as $p$ is large, our results do not significantly depend on its precise value.
In the limit $p\gg1$, the effective resistivity simplifies to
\begin{equation}
\label{eq:best}
\eta_{\text{eff}}\simeq \frac{|\mathbf{E}^*|}{e n_t c} 
\end{equation}
which has several advantages: it is simple, coordinate-agnostic, and no longer depends on the free parameters $\alpha$ and $p$, i.e., {it holds for any guide field strength}. It retains the dependence on density which we already emphasized as being of key importance.  The approximation $p\gg1$ holds for all guide fields $B_g/B_0\geq 0.3$, see \autoref{fig:p(pg)}. In \autoref{app:high_p}, we demonstrate that solutions with $p\gg1$ provide a satisfactory fit also for the case of zero guide field. We therefore regard \autoref{eq:best} as the most promising form of effective resistivity to implement in resistive MHD simulations of relativistic reconnection, {especially in global problems where it is non-trivial to determine the guide field strength.} 

We conclude with three caveats. First, our results are based on 2D simulations. While the physics of particle acceleration in relativistic reconnection is dramatically different between 2D and 3D \citep[e.g.,][]{zhang_21,zhang_23,chernoglazov_23}, the nonideal physics of field dissipation---the focus of our work---is roughly the same \citep[e.g.,][]{sironi_spitkovsky_14,werner_17}. Yet, dedicated 3D simulations should be performed to confirm our findings. Second, we have employed an electron-positron composition, and future work is needed to confirm our results in the case of electron-proton and electron-positron-proton plasmas. Finally, the generalization of our prescription to the regime of {trans-} or non-relativistic reconnection is far from trivial. In fact, the importance of charge starvation and compressibility effects in our prescriptive model, as emphasized in \autoref{subsec:guess_method}, is likely to change in the case of low magnetization. {There, the plasma beta becomes another important parameter governing the reconnection physics. We defer the investigation of the effective resistivity in trans- and non-relativistic reconnection (for different plasma beta) to future work.}

\begin{acknowledgments}
We are grateful to Fabio Bacchini, Ashley Bransgrove, Camille Granier, Rony Keppens, Oliver Porth, Sasha Philippov and Eliot Quataert for useful discussions. L.S. acknowledges support from DoE Early Career Award DE-SC0023015, NASA ATP 80NSSC24K1238, NASA ATP 80NSSC24K1826, and NSF AST-2307202. This work was supported by a grant from the Simons Foundation (MP-SCMPS-00001470) to L.S. and B.R., and facilitated by the Multimessenger Plasma Physics Center (MPPC), grant PHY-2206609 to L.S. and S.S.. B.R. \ and A.L. are supported by the Natural Sciences \& Engineering Research Council of Canada (NSERC). B.R. is supported by the Canadian Space Agency (23JWGO2A01). B.R. acknowledges a guest researcher position at the Flatiron Institute, supported by the Simons Foundation.
ERM acknowledges support by the National Science Foundation under grants No.
PHY-2309210 and AST-2307394, and from NASA's ATP program under grant
80NSSC24K1229. The computational resources and services used in this work were partially provided by Columbia University (Ginsburg HPC cluster) and by facilities supported by the Scientific Computing Core at the Flatiron Institute, a division of the Simons Foundation.

\end{acknowledgments}

\appendix 

\section{Additional Validations}\label{app:tests}

\begin{figure*}
    \centering
    \includegraphics[width=.97\linewidth]{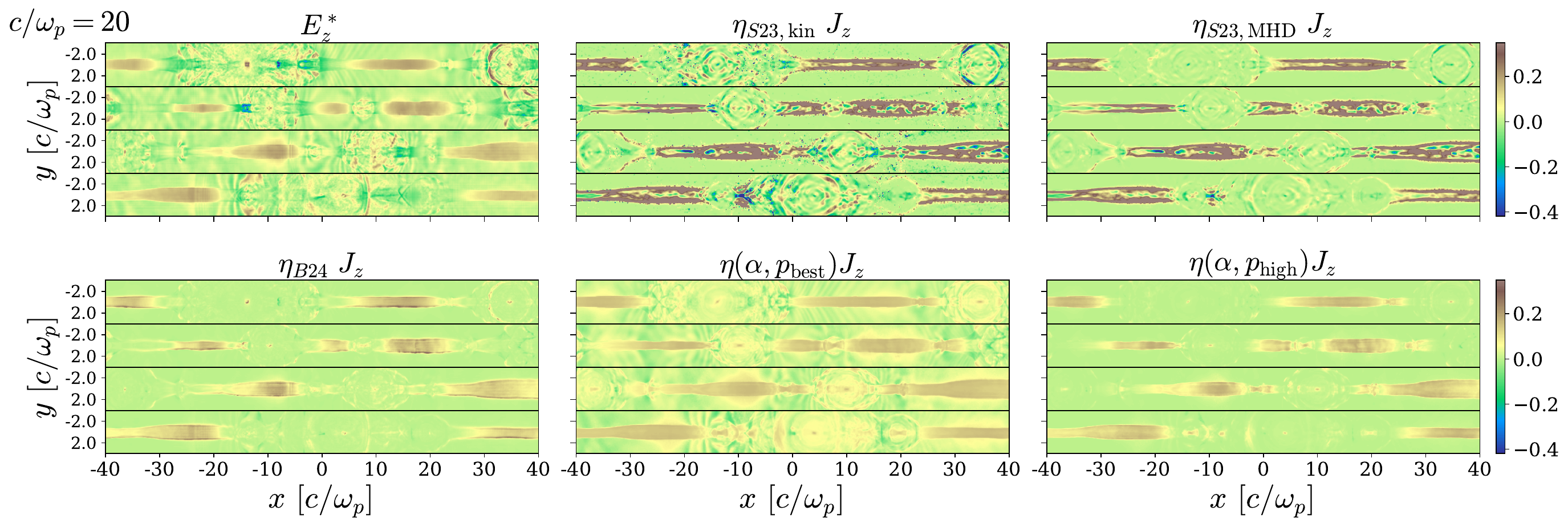}
    \caption{A comparison of the measured $E_z^*$ and various reconstructions $\etaeff J_z$ using different forms of the resistivity. Top row, from left to right: ground truth,   \autoref{Eq:Selvi}, and \autoref{Eq:SelviMHD}. Bottom row, from left to right:  \autoref{Eq:bugli} and \autoref{Eq:Guess} for both $p_{\rm best}$ and $p_{\rm high}$ (and their respective best-fit $\alpha$ values, calculated from the functions in \autoref{tab:alpha_of_p}). The simulation has zero guide field and a spatial resolution of $c/\omega_p=20$ cells. All panels are normalized to $B_0$. Within each panel, horizontal black lines separate different time snapshots: at the start of the quasi-steady phase, and after 112.5, 202.5, and 270 $\omega_p^{-1}$.
    } 
    \label{fig:hr_selvi}
\end{figure*}

\begin{figure*}[b]
    \centering
    \includegraphics[width=.97\linewidth]{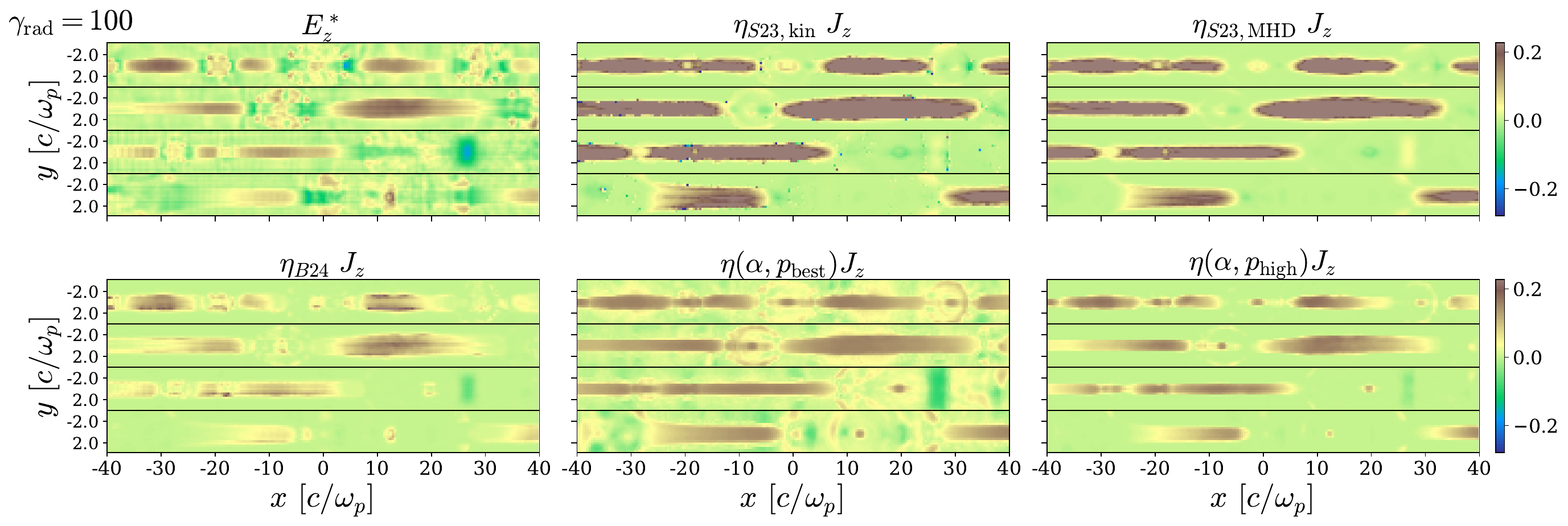}
    \caption{Same as \autoref{fig:hr_selvi}, but for standard resolution ($c/\omega_p=5$ cells) and with the addition of synchrotron cooling losses, as quantified by $\gamma_{\rm rad}=100$ (weak cooling regime, since $\gamma_{\rm rad}>\sigma$). Within each panel, horizontal black lines separate different time snapshots: at the start of the quasi-steady phase, and after 450, 810, and 1080 $\omega_p^{-1}$.
    }
    \label{fig:cooling_low}
\end{figure*}

We validate our results on two additional sets of simulations having zero guide field: first, we increase the spatial resolution, and then we perform simulations with strong synchrotron cooling. In all the cases, we find that our prescription in \autoref{Eq:Guess}---using the same $\alpha$ and $p$ as determined in the main text, see \autoref{tab:alpha_of_p}---provides a successful reconstruction of the nonideal field.

We first confirm our findings with a higher resolution simulation ($c/\omega_p$ = 20 cells) having zero guide field. The length of the domain in the $x$ direction is $L_x = 480\,c/\omega_p$. The results in \autoref{fig:hr_selvi} confirm the robustness of our conclusions, with $\etaeff(\alpha,p_{\rm high}) J_z$ visually providing the best proxy for the nonideal field.

\begin{figure*}
    \centering
    \includegraphics[width=.97\linewidth]{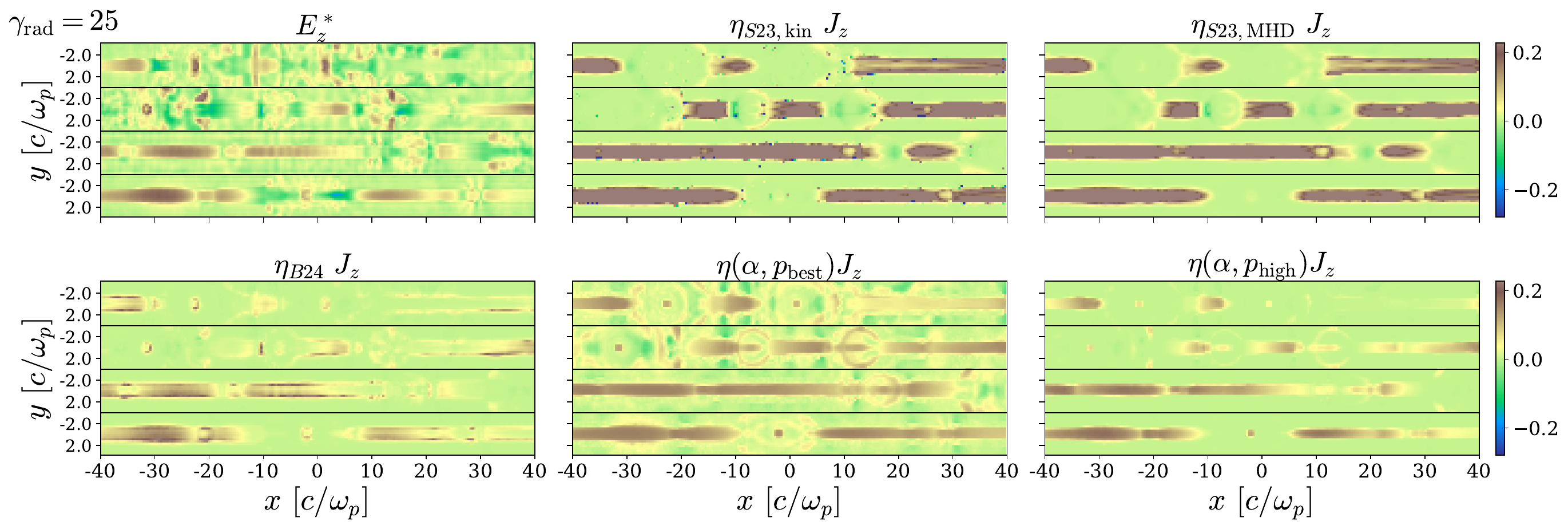}
    \caption{
    Same as \autoref{fig:hr_selvi}, but for standard resolution ($c/\omega_p=5$ cells) and with the addition of synchrotron cooling losses, as quantified by $\gamma_{\rm rad}=25$ (strong cooling regime, since $\gamma_{\rm rad}<\sigma$). We show the same time snapshots as in \autoref{fig:cooling_low}.}
    \label{fig:cooling_high}
\end{figure*}

We also perform simulations with synchrotron cooling losses and the fiducial resolution of $c/\omega_p=5$ cells. We quantify the cooling strength  via the radiation reaction Lorentz factor $\gamma_{\rm rad}$, also known as the classical  ``burnoff'' limit, at which the radiation-reaction drag force balances the accelerating force of the reconnection electric field, yielding 
\begin{equation}
    \gamma_{\rm rad}=\sqrt{\frac{e (v_{in}/c) B_{0}}{(4/3)\sigma_{\rm T}(B_0^2/8\pi)}}~.
\end{equation}
The results in \autoref{fig:cooling_low} and \autoref{fig:cooling_high} confirm the robustness of our conclusions, both for weak ($\gamma_{\rm rad}=100$) and strong ($\gamma_{\rm rad}=25$)  cooling. In particular, our prescription $\etaeff(\alpha,p_{\rm high}) J_z$ visually appears to provide the best proxy for the nonideal field. In summary, \autoref{Eq:Guess}---with $\alpha$ and $p$ determined from the fiducial simulations discussed in the main text---can be successfully applied to other runs, including the important case of strong cooling losses.

\section{The Full Ohm's Law} \label{app:full_Ohm}
\begin{figure*}
    \centering
    \includegraphics[width=0.95\linewidth]{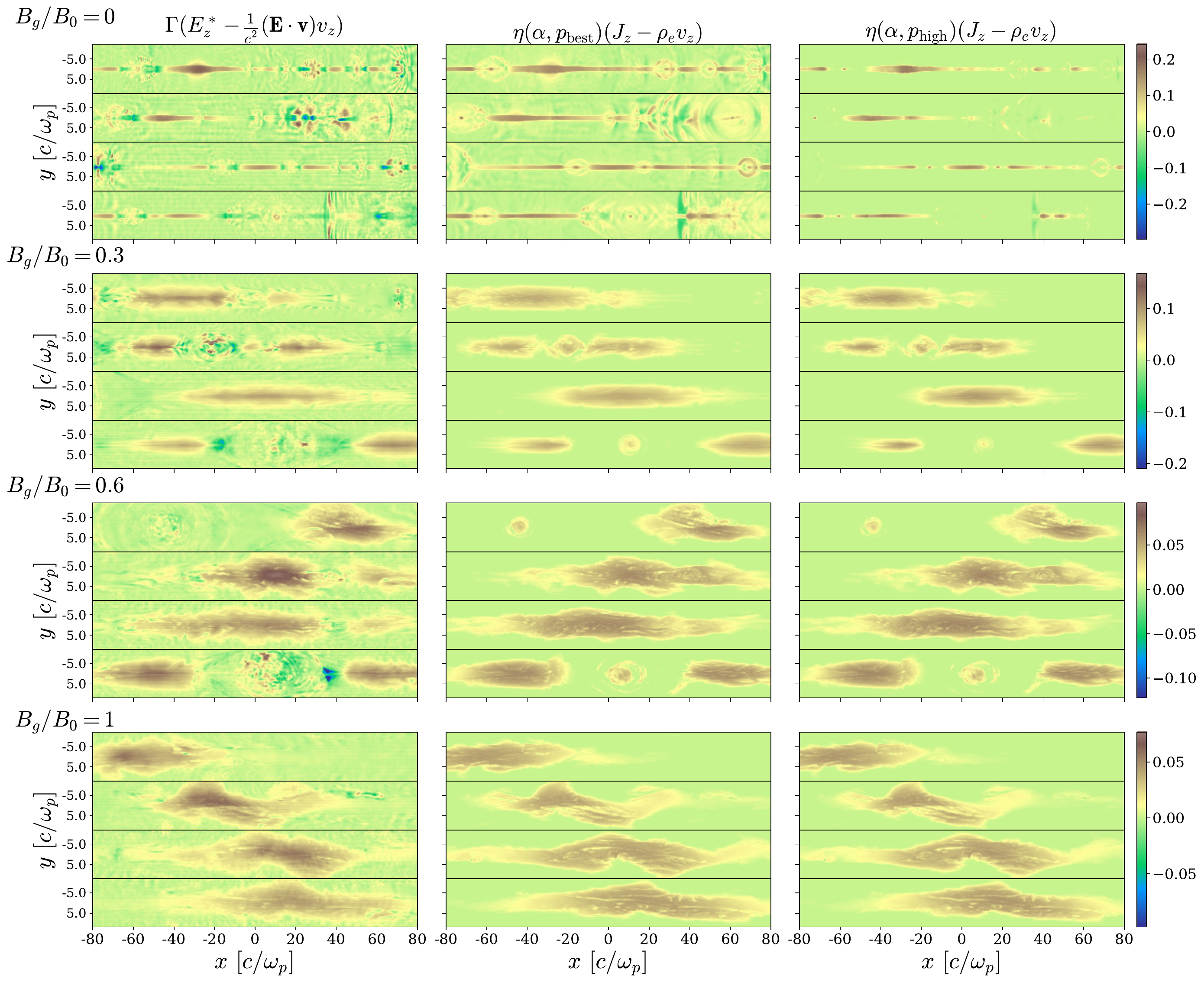}
    \caption{Same as \autoref{fig:guess_comp}, but including all terms in the Ohm's law for resistive relativistic single-fluid MHD (\autoref{eq: ohm_full}) instead of the simplified form in \autoref{Eq:eta} which we used in the main paper. We still employ our prescription (\autoref{Eq:Guess}) with the same $\alpha$ and $p$ as discussed in the main text.
}
    \label{fig:full_Ohm} 
\end{figure*}

In the main text, we reduced the full Ohm's law for resistive relativistic single-fluid MHD (\autoref{eq: ohm_full}) to the simpler form in \autoref{Eq:eta}. 
We verify in \autoref{fig:full_Ohm} that our results still hold when using the full relativistic Ohm's law for resistive MHD, as given in \autoref{eq: ohm_full}. Differences with respect to \autoref{fig:guess_comp} are minor, especially in lower guide field cases. For simulations with stronger guide fields, we see that when the full Ohm's law is used, the agreement between our model and the ground truth in plasmoid cores improves.  

\section{Extending the range of \textit{p} for zero guide field}\label{app:high_p}
\begin{figure*}
    \centering
    \includegraphics[width=0.95\linewidth]{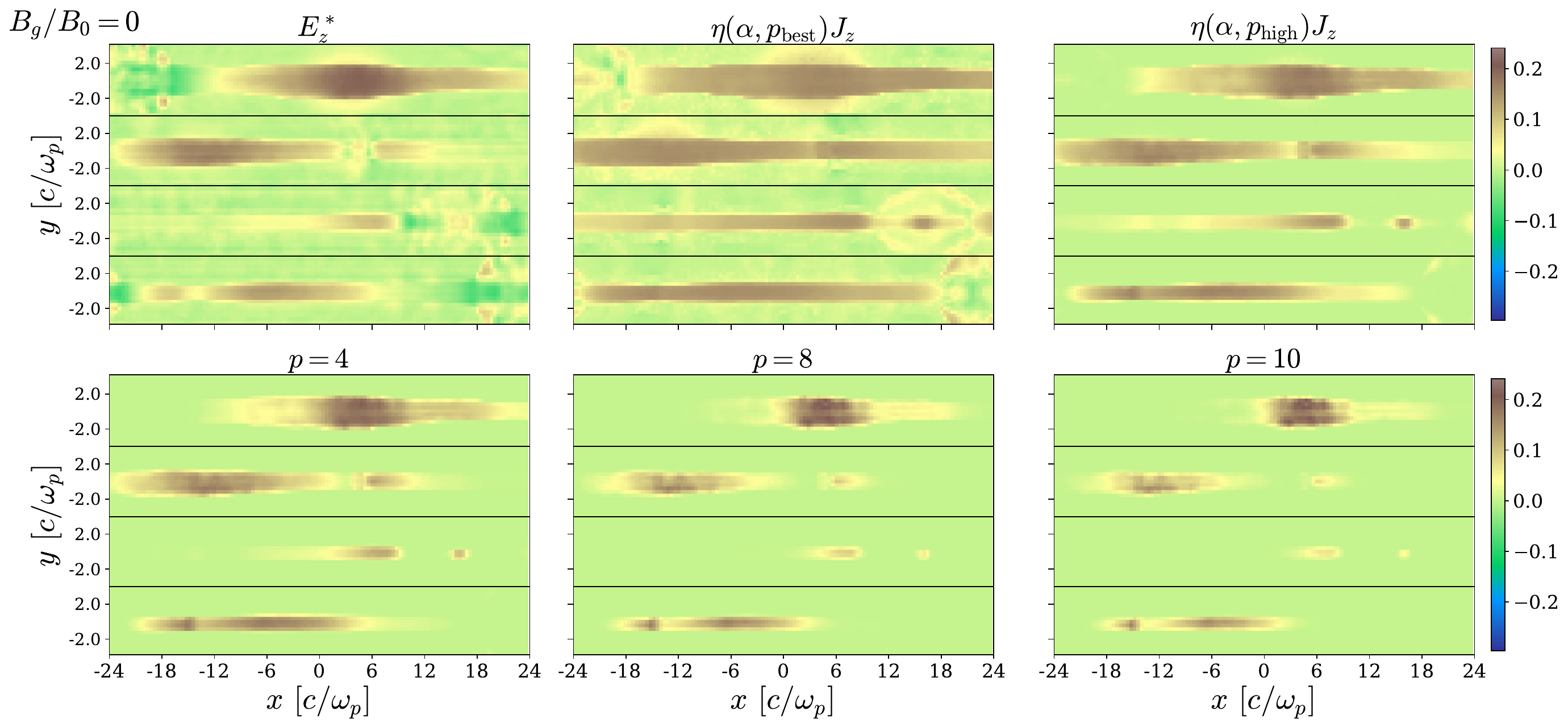}
    \caption{A comparison of the measured $E_z^*$ and the reconstruction $\etaeff J_z$ using \autoref{Eq:Guess}. We vary $p$ beyond the range given in \autoref{tab:search_params} and for each $p$ we use the optimal $\alpha$ value calculated from the functions presented in \autoref{tab:alpha_of_p}. The simulation has zero guide field, no cooling, and a spatial resolution of $c/\omega_p=5$ cells (i.e., it is the reference run used in the main paper). All panels are normalized to $B_0$. The time snapshots are the same as in \autoref{fig:selvi_comp1} and \autoref{fig:guess_comp}. }
    \label{fig:p_range_0} 
\end{figure*}

In \autoref{fig:guess_comp}, we have shown that higher values of $p$ appear to reconstruct more accurately the nonideal electric fields in the case of zero guide field, despite yielding formally higher loss values. Motivated by this, we explore how the 2D spatial structure of $\etaeff J_z$, with $\etaeff$ in  \autoref{Eq:Guess}, changes when using values of $p$ higher than $p_{\rm high}$ (for each $p$, we use the $\alpha$ value given by the function in \autoref{tab:alpha_of_p}). The results shown in \autoref{fig:p_range_0} show that values of $p$ greater than $p_{\rm high}$ up to at least $p=4$ provide an excellent reconstruction of the ground truth.

\end{document}